# The Central Role of Energy in the Urban Transition: Global Challenges for Sustainability

Joseph R. Burger[1*], James H. Brown[2], John W. Day Jr.[3], Tatiana P. Flanagan[4,5], Eric D. Roy[6]

[1]Duke University Population Research Institute (DUPRI), Gross Hall, 140 Science Drive, Durham, NC 27708 (JRB)
[2]Department of Biology, University of New Mexico, Albuquerque, NM 87106 (JHB)
[3]Department of Oceanography and Coastal Sciences, College of the Coast and Environment, Louisiana State University, Baton Rouge, LA 70803 (JWD)
[4]Department of Computer Science, University of New Mexico, Albuquerque, NM 87106 (TPF)
[5]Sandia National Laboratories, Albuquerque, NM 87101 (TPF)
[6]Rubenstein School of Environment and Natural Resources, University of Vermont, Burlington, VT 05405 (EDR)

*Corresponding author's e-mail: jrb130@duke.edu (JRB)

**Abstract**
The urban transition, the increased ratio of urban to rural population globally and within countries, is a hallmark of the 21st century. Our analysis of publicly available data from the World Bank spanning several decades for ~195 countries show that across and within nations over time, per capita Gross Domestic Product (GDP), energy use, and $CO_2$ emissions are lowest in predominantly rural countries (rural > urban pop.), increase rapidly across urbanizing countries (rural ≈ urban pop.) and are highest in the most urban countries (rural < urban pop.). These trends coincide with changes in employment by sector and gender. Rural economies are based largely on employment in the resource-extraction sector, which includes agriculture, fisheries, forestry, and mining. In urbanizing nations, male employment is predominantly in the industrial sector, including public utilities, while female employment is higher in service-based than resource-based economies. In the most urban nations, service economies predominate with some countries employing 90% of women and 65% of men in the service sector. Our analysis shows that per capita GDP, energy use, and $CO_2$ emissions increase by over two orders of magnitude from low-income, resource-based rural countries to high-income, urbanized countries with predominantly service economies. Data from the U.S. over the past 200 years illuminate a socio-metabolic urban transition similar to that seen globally in recent decades across countries and through time. Our study suggests that increased energy demand and climate consequences of burning fossil fuels will continue to accompany a rapidly urbanizing planet posing major challenges for global sustainability.

**Introduction**

The unprecedented growth of the global human economy and population is characterized by feedbacks between resource use, demography, and innovation (Boserup 1965, Bettencourt et al. 2007, Nekola et al. 2013, Weinberger et al. 2017, Hall and Klitgaard 2017, Burger 2018). A relatively recent phenomenon is the global urban transition where more people now live in cities than rural areas (Figure 1). Developing a science of cities and urbanization is a vibrant area of research that transcends the boundaries of the physical, biological, and social sciences (Seto et al. 2012, Brelsford et al. 2017, Acuto et al. 2018). Research into the drivers and consequences of urbanization is necessary to disentangle the complex interactions between the socio-economy and the biophysical environment that determine the present status and future trajectory of global population and economy.

The increased ratio of urban to rural population and the rapid increase in the number of large cities have been interpreted as encouraging signs that the human population and economy are on a path toward global sustainability. The underlying premise is that concentration of the Earth's growing population and industrial-technological-informational economy in urban centers will moderate human impacts on the environment and make for more efficient use of the planet's limited space and natural resources (Newman 2006, Jenks and Burgess 2000, Glaeser 2011). Recent research on scaling of relevant variables with city size suggests that increasing returns in Gross Domestic Product (GDP) and innovation, including research, development, and "super creative" activities, can be attributed to the enhanced social networks facilitated by the "buzz of the city" (Bettencourt et al. 2007, Bettencourt and West 2011, Bettencourt 2013). High-density housing and efficient transport systems for goods, people, and services (Burton 2000, Capello and Camagni 2000) lead to economies of scale in space use and infrastructure (Bettencourt et al. 2007). Indeed, the last few centuries have been marked by consistent gains in multiple indicators of quality of life (Brown et al. 2014, Steffen et al. 2015). However, much of these gains are directly or indirectly due to increased overall and per capita energy use (Lambert et al. 2014).

These trends are the basis for optimism that continued urbanization and socioeconomic development will lead to reduced poverty, new educational and employment opportunities, and increased quality of life for billions of people. Since all developed nations are predominantly urban (Satterthwaite et al. 2010), urbanization is often viewed as an important contributor to a sustainable future (Glaser 2011). Prediction of a future trajectory without scientific grounding and based primarily on current trends is hazardous, however, because it ignores fundamental biophysical and socioeconomic constraints on the relation between humans and their environment. In particular, the increase in global human population and economy has been accompanied by increased extraction and use of natural resources, including extra-metabolic energy largely in the form of fossil fuels (Brown et al. 2011, Burger et al. 2012, Burger et al. 2017). To assess the sustainability of urbanization and its correlates, and consequences will require quantitative analyses of the energy and resource flows that fuel the connection of urban systems to rural and global systems across large space and time scales.

Many urban studies assume or imply that urbanization can decouple economic growth from environmental impacts by concentrating population and technology in efficient and innovative cities. It has also been suggested that dematerialization accompanies the movement to a service economy (Romm 2002, Victor 2010). This is incorrect. As socio-economic systems grow, the infrastructure to support its organization, and the flow of resources and information must also grow. Urban systems use fossil fuels to increase the flux of vast quantities of resources across urban boundaries in order to keep hyper-dense populations and their lifestyles alive (Rees and Wackernagel 1996, Rees 2012, Burger et al. 2012, Burger et al. 2017). It is also incontrovertible that socioeconomic trends at all scales—from local to global—are subject to biophysical laws that govern relations between humans and their environments (Daly 2005, Burger et al. 2012, Day et al. 2016, 2018, Hall 2017). Human resource consumption has caused climate change and environmental degradation that feeds back to pervasively affect demography and economies from local (Burger et al. 2012) to global scales (Brown et al. 2014). Energetic accounting approaches allow the quantification of the increased energy subsidies that have led to the transition of humans from hunter-gatherers, to agriculturalists, to contemporary societies with dense concentrations of people and socioeconomic activities in large cities (Krausman et al. 2008, Burger et al. 2017). A recent analysis of household carbon footprints in the U.S. found no evidence that increasing population density in urban cores or suburbs creates net greenhouse gas emissions benefits when considering entire metropolitan areas (Jones and Kammen 2014). In China, urban households produce more than twice as much $CO_2$ per capita as rural households, resulting in the increased carbon emissions that have accompanied China's urban transition (Feng and Hubacek 2016). Scaling of energy use and $CO_2$ emissions with city size in the U.S., China, and Europe is either linear (directly proportional to population) or super-linear (larger per capita emissions in larger cities; Bettencourt et al. 2007, Fragkias et al. 2013, Oliveira et al. 2014). However, a general understanding of the resource, economic, and demographic correlates and consequences of the urban transition globally is still lacking.

In this study, we focus on the role of energy in the urbanization process and its interrelationships with climate, economic, and demographic metrics. Our objectives are to: i) present a conceptual socio-metabolic model of urbanization linking the interrelationships among biophysical and social variables, ii) document empirical trends in per capita energy use, $CO_2$ emissions, GDP and employment by economic sectors across countries and within countries over time, and iii) statistically determine change-points in the trajectories in these variables in relation to the urban transition.

**The socio-metabolism of urbanization**

We developed a simplified conceptual framework to begin to disentangle the important variables and complex feedbacks between the biophysical environment and the human socio-economy that are most relevant to modern human ecology and urbanization. This socio-metabolic perspective follows Fischer-Kowalski and Haberl (2007), Krausmann et al. (2008) and others by taking into account the central role of energy transitions that have allowed the relatively continuous growth of the socio-economy over human history. Specifically, it highlights the central role of increasing consumption of resources from the biophysical environment – especially energy – to fuel urbanization, grow the economy, and shift from resource extraction-based sectors in rural areas, to industrial and service economies in cities (Figure 2).

This model is based on fundamentals of energetics in ecology and society (Hall and Klitgaard 2018). Like all complex biological systems, exchanges of energy and materials with the environment are necessary for growth, development, and maintenance of the human population and economy. This results in inputs of energy and raw material resources (harvest and production in Figure 2) and outputs of people, manufactured products, and wastes. As societies develop, shifts from hunter-gatherer to agrarian to industrial-technological-service economies are accompanied by increased consumption of energy and materials to support higher population densities, greater societal complexity, and increased economic productivity (Krausman et al. 2008, Day et al. 2016).

The model portrays human society comprised of rural and urban populations embedded in the biophysical environment. In the model, there are differences between rural resource-based societies with low energy input, and urban industrial-technological-service societies with high extra-metabolic input from fossil fuels. In pre-industrial societies, rural populations were much larger than urban populations (Figure 2). They used human labor, draft animals, and relatively simple technologies (e.g., plows, axes) to harvest food (for biological metabolism) and other raw materials (e.g., Fizaine and Court 2016). As socio-economic development occurred, rural populations were able to acquire surplus food and raw materials (wood, fiber, minerals, skins, etc.) to trade with growing urban populations for manufactured goods and services.

In modern industrial societies, urban populations are much greater than rural populations, and flows of resources from remote countryside, rivers, coasts, and oceans to cities have increased dramatically, due largely to energy subsidies mainly in the form of fossil fuels. Use of coal and then oil and gas have increased enormously since the industrial revolution (Smil 2008). They currently account for over 80% of total human energy use with concomitant increased $CO_2$ emissions. Urban populations are now the dominant contributor, directly and indirectly, to global GDP, use of energy and other natural resources, $CO_2$ emissions, and climate change (Acuto et al. 2018). The contemporary global socioeconomic system relies on exchanges with the biophysical environment and flows (trade) between rural areas and urban centers for its continued functioning, maintenance, and growth (Burger et al. 2012). Extraction of natural resources in rural areas and flows into urban areas (trade exchanges) with subsequent release of waste back to the environment at local to global scales form the bases of the modern industrial-technological-service economy. The logical links among biophysical, demographic, and economic variables in Figure 2 motivated our analyses of data across countries and years.

**Methods**
*Data*

We used publically available World Bank data for per capita energy use, $CO_2$ emissions, Gross Domestic Product, and energy/GDP intensity to quantify their relationships with urbanization. The World Bank data are aggregated for geographic regions (e.g., Latin America), development status, and includes non-country entities (e.g., Guam, Hong Kong SAR, China). So, we removed all non-country rows in The World Bank data resulting in ~195 countries spanning up to 50 years depending on variable. The World Bank development indicators are regularly updated, so we refer the reader to https://data.worldbank.org for the most recent version.

We quantified urbanization as the $\log_{10}$ ratio of rural to urban population within a country. We use a $\log_{10}$ transformation because raw ratios span more than two orders of magnitude (Brown et al. 2011, 2014). We use $\log_{10}$ transformed country level data on per capita energy use (thousand tonnes of oil equivalent: KTOE), $CO_2$ emissions (tonnes), Gross Domestic Product (GDP, in constant US$), and energy intensity (defined as per capita energy use divided by per capita GDP) to quantify their relationships with urbanization. We used GDP in constant US$ because it is adjusted for inflation, available for all countries spanning decades, and highly correlated with other measures of economic well-being, including the Human Development Index (Brown et al. 2014). One limitation of these data is that energy use and $CO_2$ emissions per capita reflect only rates within the borders of a country (i.e., domestic fuel combustion) and do not include the total energy used to support a country and its GDP (Brown et al. 2011). For example, CO$_2$ emissions resulting from the production of goods in China that are exported to consumers in the United States are attributed to China and not to the US (Weber et al. 2008). Consequently, these estimates of energy use and CO$_2$ emissions are under-estimated for more developed countries and over-estimated for less developed countries.

We examined changes in employment among resource, industrial, and service sectors across the urban transition allowing us to gain insights into societal changes that link the biophysical environment to economic productivity. *Resource employment* is defined as the total number of individuals in a given country that are publicly or privately employed in resource-extraction occupations, including agriculture, hunting, forestry, and fishing. *Industry employment* is the total number of individuals that work in industrial jobs, including mining and quarrying, oil and gas production, manufacturing, construction, and public utilities such as water, electricity, and, gas. *Service employment* is the total number of individuals employed in the service economy, including wholesale and retail trade, hotels and restaurants, transportation, storage, communications, finance, insurance, real estate, business services, health care, education, recreation, and community, social, and personal services.

*Analysis*

To initially visualize the relationship between urbanization and per capita energy consumption, CO$_2$ emissions, and economic growth, we fit splines through all data points for all countries and years. Splines are a flexible data-driven approach that allows easy examination of the general patterns of each variable with urbanization across countries and years. Although data with large standard deviations may allow for simpler models to calculate regressions, spline analysis is appropriate when plots of data show visual non-linear changes in structure (e.g., Hurley, 2004), as is the case in our study. Spline analysis uses a smoothing parameter, $\lambda$, which balances the tradeoff between fidelity of the data and roughness of the function estimate by penalizing any curvature. As $\lambda \to \infty$, only linear functions are allowed since any curvature at all becomes penalized with an infinitely large number. A $\lambda$ value of 0 (no smoothing) means that the smoothing spline converges to an interpolated spline fitting all data points. We used the default value of $\lambda = 1$ in the programming software Matlab (2015) which showed a balance between data fidelity and roughness of the curve.

Since plots of the data clearly showed non-linear changes along the rural-to-urban continuum, we used a classical analysis to determine statistically the change-points using cumulative sums. Analyzing the cumulative sum of residuals was first proposed by Brown et. al (1975) as a technique to detect significant departures from a regression. This technique allows detection of change-points by calculating the cumulative sum of the differences between individual data values and the mean of the data. If the data do not deviate from the mean trend, the linear plot of these differences does not show pronounced changes in slope and the range (the difference between the highest and lowest points) is small. In data trends with substantial shift, cumulative sums show a visible change at the point where the change-point occurred. We detected only one significant change-point for each variable along the urban transition. After calculating a first level change-point for each dependent variable as a function of urbanization (the log$_{10}$ of the ratio of urban to rural population), we used a regression (Ordinary Least Squares, OLS) to quantify the changes in each subset of the data, before and after the change-point, and compared the resulting regression model parameters (slopes and intercepts and their 95% confidence intervals).

**Results**

Increased urbanization was associated with higher national per capita energy consumption, $CO_2$ emissions and economic growth (i.e., GDP; Figure 3). Per capita energy use was lowest in countries with predominantly rural populations and near-subsistence economies: per capita GDP was usually less than 1,000 US$ per year and energy use less than 500 watts, or about five times the human metabolic rate of about 100 watts (~2000 calories/day). In the most developed countries with predominantly urban populations, per capita GDP and energy expenditure use was much higher. GDP per capita was usually greater than 50,000 US$ per year, and energy use was typically more than 5,000 watts, so more than 50 times the human metabolic rate (Figure 3).

Fitting splines to the data showed that there were statistically significant socio-metabolic shifts associated with urbanization (Figures 3, 4). Per capita energy use and $CO_2$ production increased slowly with changes in urbanization across the predominantly rural countries, then was steeper through the urban transition (~ 50% urban), and then began to level off in the predominantly urban countries (Table 1). Although $CO_2$ emissions per capita showed signs of leveling off in the most urbanized countries, these levels were much greater (approximately two orders of magnitude) than in the most rural countries. GDP, by contrast, increased at a generally similar rate across the rural-to-urban spectrum, but more steeply in the most urbanized countries (Figure 4). In the most urbanized countries, the splines show decreasing rates of growth in per capita energy use and $CO_2$ production and a leveling off of GDP per capita. However, there are few countries to the far right of the graph so there is greater uncertainty.

We detected one change-point in the data across the rural-to-urban gradient for each variable, and all of these occurred around the urban transition (50/50). OLS regression models fitted to data before and after the change-point revealed that energy use and $CO_2$ emissions per capita had significantly steeper slopes in predominantly rural countries. In contrast, GDP had a steeper slope in countries to the right of the change-point with a greater ratio of urban population (Table 1, Figure 4), suggesting a shift to greater economic productivity in urbanized countries. In contrast, per capita energy use and $CO_2$ emissions show steeper slopes in rural countries, and shallower slopes in the urbanized countries. This suggests that the urban countries use both greater energy use, per capita, but also show increasing energy use efficiency in comparison to rural countries (Table 1; Figure 4). For $\log_{10}$ energy use/GDP per capita there was no statistical difference in the slopes before or after the change-point (Table 1).

The data show substantial changes across the urban transition in employment by sector and gender (Figure 5). In the most rural countries, employment was predominately in occupations related to resource extraction, such as agriculture, timber, and fisheries. Industrial employment increased with urbanization across predominantly rural countries, peaked at approximately the urban transition (50% urban) and then declined in predominantly urban countries. Service employment was dominant in the most urban countries, accounting for approximately 90% of employed women and 65% of employed men. The pervasive trend in employment across the rural-to-urban gradient was a decline in the proportion of the population employed in resource sectors and increase in the proportion working in non-resource (i.e., industrial and service) sectors, especially in the service sector in the richest countries.

**Discussion**

Our study shows how large increases in per capita energy use have fueled the rapid growth of cities and economic productivity in urbanized countries. Over decades, across countries and within countries over time, increased urbanization accompanied increases in per capita energy use, $CO_2$ emissions, and GDP. Employment decreased in resource-based economies and increased in industrial and service economies in the most urban countries. Fossil fuel supplements in predominantly rural, resource sectors of the economy also subsidize the production of basic commodities (food, wood products, minerals) that contribute to per capita productivity of non-resource based, urban economic sectors. The use of fossil fuels for modern industrialized agriculture, fisheries, logging, and mining in rural areas has substantially increased the rate of resource flow to urban areas, allowing the growth of industry and service employment that have contributed to increased urban population and economic growth.

As societies develop and populations of countries transition from rural to urban, there has been a pervasive trend of dramatically increasing per capita energy use. The great majority of this is extra-metabolic energy obtained primarily by extracting and burning fossil fuels (Smil 2008, Brown et al. 2011, Hall 2017, Day et al. 2018). In countries with predominantly rural populations and near-subsistence economies, per capita energy expenditure is only a few times more than the human metabolic rate of about 100 watts. This reflects a heavy reliance on human and animal labor and the limited use of fossil fuel and electricity to power machines. In comparison, the most developed countries with predominantly urban populations have per capita energy expenditures and $CO_2$ emissions about two orders of magnitude higher, exceeding 10,000 watts in nations such as the U.S., U.K., Norway, and Canada (Brown et al. 2011, 2014).

The increase in energy use across the urban transition reflects complex interrelated socioeconomic changes. In part, as evidenced by concomitant increases in GDP, these are correlates and consequences of economic growth and development (Poumanyvong and Kaneko 2010, Brown et al. 2011) and shifts in employment from resource extraction to industrial and service economies. The least developed countries with predominantly rural populations have consistently low rates of per capita energy use. Their economies are largely based on near-subsistence agriculture for domestic food consumption and extraction of natural resources for export to more developed countries. Thus, the extraction of natural resources for export mainly benefits developed countries. By contrast, the most developed countries with predominantly urban populations have high rates of per capita energy use. Despite recent increases in renewable energy sources, urban economies still directly and indirectly depend heavily on fossil fuels to power energy-demanding industries and services, and to build and maintain complex infrastructure networks for transportation, trade, and communication.

The increase in energy use across the urban transition is also in part a consequence of impacts of cities on rural areas that are often far away. Cities are dependent on very large flows of energy and materials – people, resources, information, and wastes – into, within, and out of urban centers. Cities require massive inputs of food, extra-metabolic energy, and other raw materials that are produced in rural areas (Rees 2012, Burger et al. 2012, 2017). They also require extensive complex transportation networks and associated infrastructure, energy expenditures and $CO_2$ emissions to move, store, and transform raw materials into goods and services (Hammond et al. 2015, Day et al. 2016). For example, transportation currently accounts for about 25% of $CO_2$ emissions globally and about 30% in the U.S. (IEA 2009). Extensive global trade networks powered mostly by fossil fuels are now necessary to maintain the flow of materials, people, information, and money and the associated regulatory, service, and information industries to sustain economic activity in urbanized areas. In recent decades, the industrial and service sectors account for 97% of global GDP and approximately 65% of global employment (Satterthwaite et al. 2010). This further highlights the dependence of service economies in developed countries and urban areas on energy and machines to supplement human labor to power economic growth.

The above two phenomena – energy expenditure in cities to support industrial-technological-informational economies and dense populations, and energy expenditure in rural areas to supply the food, shelter, fossil fuels, minerals, and other materials to support these activities in cities – are necessarily somewhat confounded. In some cases, developing countries differ in the degree of urbanization. For example, Afghanistan, Burundi, and Rwanda are still predominantly rural and have economies based mostly on subsistence agriculture and extraction and export of raw materials. Other countries, such as Mexico, the Philippines, and Bangladesh, have been urbanizing rapidly as populations have migrated to cities, even though economic benefits of urbanization have been slower to materialize as evident by low GDPs and energy use (Brown et al. 2011, 2014). There are also wide differences among highly urbanized countries. For example, Singapore, Switzerland, and Japan import most of their food, energy, and other raw materials. Others such as the U.S., Canada, and Australia have energy intensive extraction sectors in extensive rural areas, despite the concentration of their populations in large urban areas. These countries produce food and other raw materials in excess of domestic requirements and export products of agriculture, fisheries, forestry, and mining to countries with deficits in resource production and other trade benefits.

Our analysis does not capture the changes in international flows of energy and resources in relation to domestic (within-country) urbanization. The resource inputs that sustain cities are supplied both by domestic production, internal transportation, and international trade. Cities import resources from fields, forests, oceans, wells, and mines far beyond national boundaries. This explains in part why GDP increases more steeply than energy use or $CO_2$ emissions across the urban transition (Table 1, Figure 4). The data we use from the World Bank are for energy consumed and $CO_2$ emitted from within a given country. However, the most urban countries with large economies import resources internationally, contributing to economic growth and hence the steep increase in GDP. Day et al. (2016) reported that the same trend applied across U.S. states. States with low per capita energy use and $CO_2$ generation were highly dependent on imports from states with high per capita energy use and $CO_2$ emissions because of the concentration of heavy industries such as fossil fuel production and refining, coal mining, and petrochemical industries in addition to industrialized agriculture, fisheries, and forestry. Our results suggest that energy and raw materials imported from developing, predominately rural countries have subsidized population and economic growth in developed urban countries. Future studies should quantify the direct and indirect interdependencies in $CO_2$ emissions and other flows at scales from households to countries to the globe (e.g., Jones and Kammen 2011). These studies can build on prior ecological footprint analyses that have studied and articulated overshoot, ecological deficits, inter-regional subsidies, and unsustainable interdependencies of modern cities (e.g., Rees and Wackernagel 1996, Rees 1997, Folke et al. 1997, Warren-Rhodes and Koenig 2001, Moore et al. 2013, Baabou et al. 2017, Isman et al. 2018). For example, Rees (2012) argues that modern industrial societies and cities are inherently unsustainable from an energy standpoint because "cities are self-organizing far-from-equilibrium dissipative structures whose self-organization is utterly dependent on access to abundant energy and material resources". This raises important global problems for sustaining current, much less future, levels of urbanization.

*Summary and challenges for the future*

This study highlights the central role of energy in the urban transition and the associated flows of $CO_2$ emissions, economic growth, and demographic changes in employment. The global urban transition is marked by vast increases in the use of extra-metabolic energy, largely in the form of fossil fuels, to supplement human and animal labour to fuel the shift from rural resource-based economies to urban industrial and service-based economies. Data from the U.S. over the past 200 years exemplify these trends (Figure 6 in Box 1). As populations became concentrated in urban centers, they increasingly relied on primary resource sectors in non-urban areas, often in other countries, to supply basic necessities such as food, water, and materials required to sustain the industrial-technological-service economies of cities. Additionally, vast quantities of energy and other resources are required to build and maintain the infrastructure necessary to support rapid and complex transportation and information networks linking cities and rural areas across the globe.

This increasing dependence of rapidly growing cities on resources imported from beyond their borders, and the infrastructure that enables this process powered by fossil fuel networks means that cities, as we know them, are inherently unsustainable (e.g., Rees 2012 quoted above). Some of these resources are non-renewable fossil fuels and metals. Others such as food from agriculture and fisheries and materials from forestry are renewable but are currently being harvested at rates that often exceed sustainable production using vast amounts of fossil energy (Burger et al. 2012). These ecological limits are consequences of inviolate biophysical laws that cannot be overturned or circumvented by technological innovations. For example, every human being requires at least 2,000 kcal of food energy and 3-4 liters of fresh water per day just to stay alive. Every city dweller depends on large quantities of imported extra-metabolic energy and raw materials often from global scale trade networks (Rees 2012, Burger et al. 2017). Non-carbon based energy sources supply only a small fraction (<15%) of total energy used by the modern economy (Brown et al. 2014, Hall 2017). Moreover, some important uses of fossil fuels will be difficult to replace with renewables including aviation, container ships, industrial machines, heavy industry, and industrial feedstocks and agriculture because the energy return on energy invested for renewable energy sources is often too low to allow complete substitution for the power afforded by high density, fossil fuel energy (Hall and Day 2009, Hall 2017, Day et al. 2018).

Our results and the inescapable biophysical realities pointed out by other studies (e.g., Daly 2005, Hall and Day 2009, Burger et al. 2012, Hall and Klitgaard 2017, DeLong et al. 2013, Brown et al. 2014, Day et al. 2016, 2018) beg questions of how much longer large urban populations and economies will be able to grow and prosper and what will happen as the inevitable resource scarcities begin to take effect? Eighty percent of the North American population is currently urban and the UN predicts that by 2050, two-thirds of the world's population will live in cities with greatest urbanization occurring in Asia and Africa (https://esa.un.org/unpd/wup/publications/files/wup2014-highlights.pdf). Our study suggests that, for these predictions to take place will require considerably more than the 18 terawatts (1 terawatt = 1 trillion watts) of annual energy throughput that maintains current trends in global urbanization. We conclude that the rapidly growing urban systems across the globe are inherently unsustainable in their current form due to the primary dependence on non-renewable resources in the form of fossil fuels that result in growing climate impacts and environmental degradation (Day et al. 2018).

Several challenges will have to be faced in an increasingly urbanized and resource-limited world. Any movement towards real sustainability must consider these biophysical realities and take concrete steps to address the root causes of non-sustainability. Our study can inform policy decisions necessary to lead us towards a cultural norm of climate stewardship. It is clear that we must reduce fossil fuel use globally. Efficiency is not the answer. Increasing efficiency does not guarantee reduced consumption (i.e., Jevon's paradox) as our results suggest. Meeting climate targets will require de-carbonization of the existing global economy at a rate (>4%/yr) that only a handful of countries (e.g., Sweden and France) have been close to historically, and this has been done largely through exporting energy intensive activities to developing countries (Hubacek et al. 2017). This does not decrease the net global $CO_2$ levels. Additional energy will be needed to lift billions of people out of rural poverty and to grow cities. We agree with others that income distribution and the carbon intensity of lifestyles needs to be a primary focus of discourse moving forward given the fact that the top 10% income earners globally (most of whom are urban-dwellers) are responsible for more than one third of the current carbon footprint of households (Hubacek et al. 2017) Therefore, we recommend policy strategies that aim to reduce fossil fuel usage by top income earners. See, for example, the National system for a "fee-and-dividend" proposed in Brundtland et al. (2012). This policy proposes a single fee levied on carbon at the point where the resource is traded on commercial markets. For example, the domestic wellhead, other production points such as mines, or ports of entry for importers. This fee is reflected in energy cost to consumers, but instead of going to government taxes, 100% of it is refunded to citizens as a dividend. The carbon fee would be passed on to consumers allowing market economies to establish price through supply and demand. There is an incentive to use less fuel. Those who use less fossil fuels, per capita, than the average will thus receive more in dividends than they would pay as prices increase. Such policies would additionally begin to reduce inequalities that lead to social tensions and unrest, which may be more proximate threats to social-economic-environmental collapse (Wilkinson and Pickett 2009, Nafeez 2017, Heinberg and Crownshaw 2018).

There is a myriad of additional solutions to reduce the carbon intensity of human activities associated with the support of urban centers, and a paradigm shift of global economic systems within biophysical realities must be advanced in parallel (Hawken 2017). In this pursuit, we need an economic system refocused on quality of life and health of the Earth, ecosystems, and biodiversity for current and future generations to guide human endeavors. This will require a fundamental reordering of economic systems based not on a growth paradigm measured as % change in GDP, but to an economic system that acknowledges the human socio-economic system as a subsystem of the biosphere and places emphasis on well-being rather than consumption (Daly and Farley 2004, Daly 2005, 2015). We should be preparing for a soft landing in a post fossil fuel world – a "prosperous way down" (Odum and Odum 2008) – now, rather than encouraging energy draining urban development that is inherently unsustainable. Time is of the essence.

**Acknowledgements**
We thank members of the Human Macroecology Group at UNM for helpful discussion and Felisa Smith, Melanie Moses, Bruce Milne, and Mathew Moerschbaecher for feedback on earlier drafts. Travis Knowles brought our attention to Fee-and Dividend Policy. Partial support for JWD was from the Gulf Research Program of the National Academies of Sciences, Engineering, and Medicine Award Number 2000005991. TPF was supported by a Postdoctoral Fellowship by a McDonnell Foundation Complex Systems Scholar Award to the Moses Lab at UNM. Sandia National Laboratories is a multimission laboratory managed and operated by National Technology & Engineering Solutions of Sandia, LLC, a wholly owned subsidiary of Honeywell International Inc., for the U.S. Department of Energy's National Nuclear Security Administration under contract DE-NA0003525. The views expressed in the article do not necessarily represent the views of the U.S. Department of Energy or the United States Government.

Conflict of interests: On behalf of all authors, the corresponding author states that there is no conflict of interest.

| Variable | | Intercept [CIs] | Slope [CIs] | R-squared | p-value |
|---|---|---|---|---|---|
| log Energy Use (KTOE) | Before CP | 2.96 [2.94, 2.97] | 0.72 [0.68, 0.75] | 0.3320 | < 0.001 |
| | After CP | 3.20 [3.17, 3.23] | 0.40 [0.35, 0.45] | 0.0941 | < 0.001 |
| log $CO_2$ (Tonnes) | Before CP | 0.14 [0.11, 0.16] | 1.22 [1.17, 1.27] | 0.3650 | < 0.001 |
| | After CP | 0.41 [0.39, 0.43] | 0.62 [0.57, 0.66] | 0.1690 | < 0.001 |
| log GDP (Constant US$) | Before CP | 3.05 [3.02, 3.07] | 0.77 [0.72, 0.81] | 0.2120 | < 0.001 |
| | After CP | 3.31 [3.29, 3.34] | 0.88 [0.83, 0.92] | 0.2650 | < 0.001 |
| log Energy Use/ GDP | Before CP | -0.26 [-0.29, -0.22] | -0.42 [-0.49, -0.34] | 0.0716 | < 0.001 |
| | After CP | -0.27 [-0.29, -0.25] | -0.35 [-0.39, -0.30] | 0.0719 | < 0.001 |

Table 1. Summary statistics for linear models before (left) and after (right) the change-point along the urban transition in Fig 4.

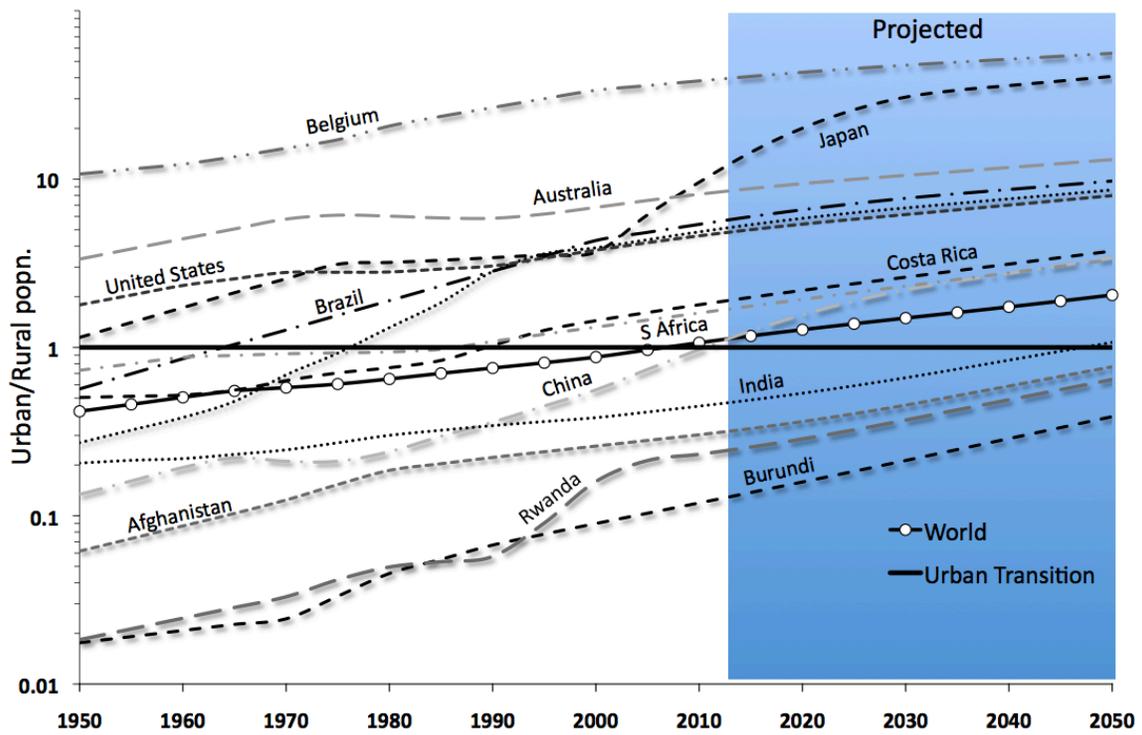

Figure 1: The urban transition globally and for select countries. Data are from the World Bank (http://data.worldbank.org).

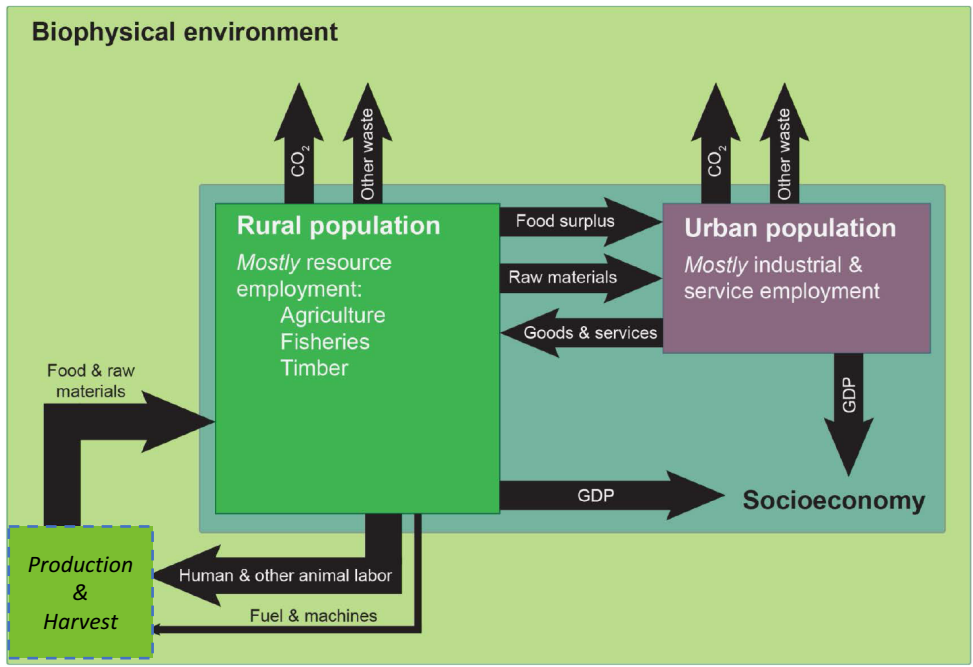

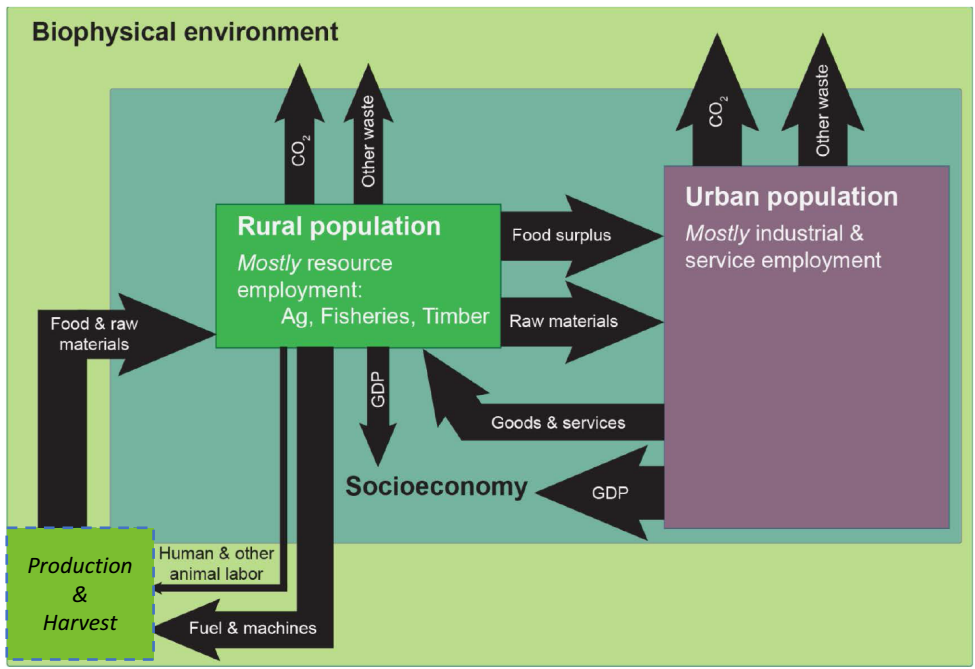

Figure 2: A conceptual model that links the human economy to the biophysical system between rural (A) and (B) urban societies through demographic changes in resource use, employment by sectors, and Gross Domestic Product. Harvest and Production refers to those components of the Biophysical Environment that are directly harvested.

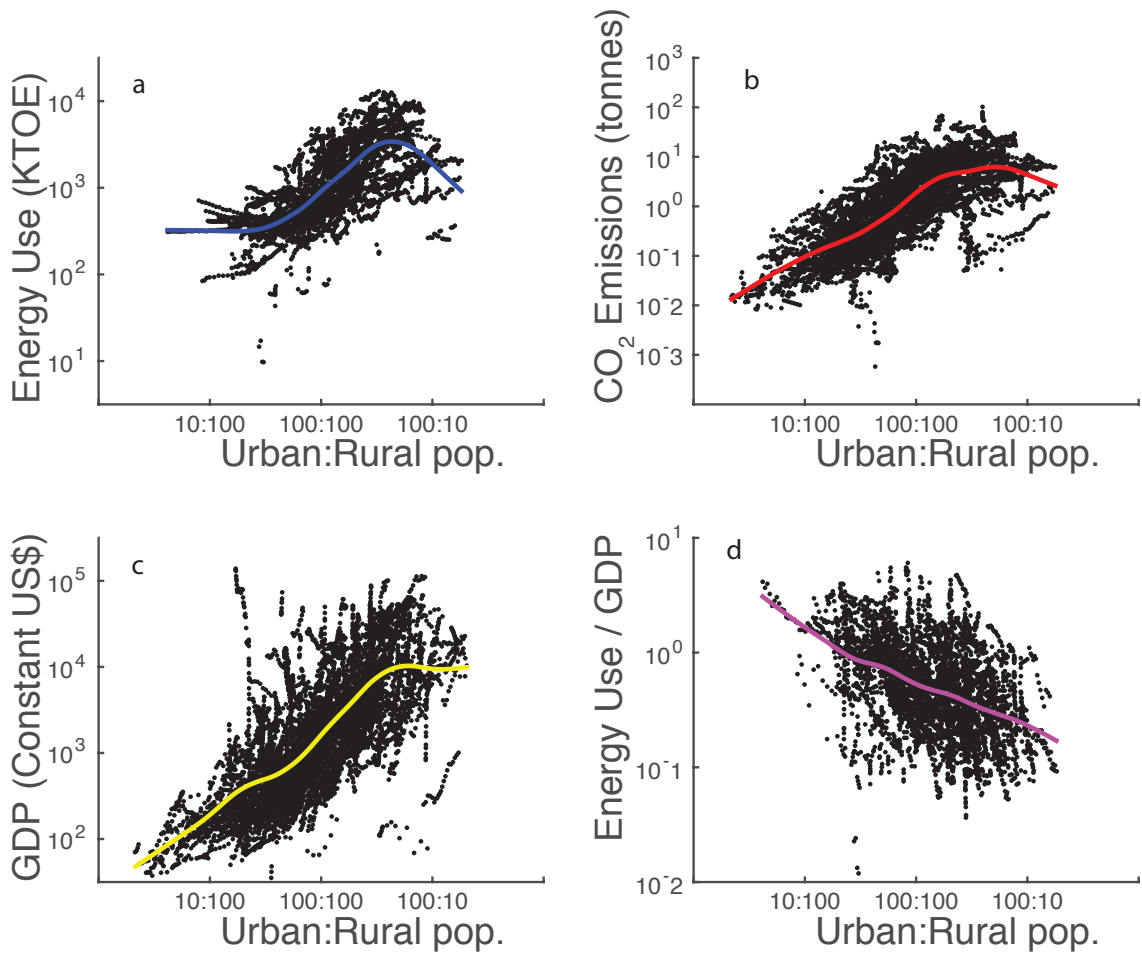

Figure 3: The relationship between log values of per capita energy use, $CO_2$ emissions, GDP, and energy use/GDP with urbanization (urban to rural population ratio) and across years and countries. Urban countries are to the right and rural countries to the left. Colored lines show smooth splines through all points irrespective of country or year. These data are from the World Bank (see http://data.worldbank.org for recent updates).

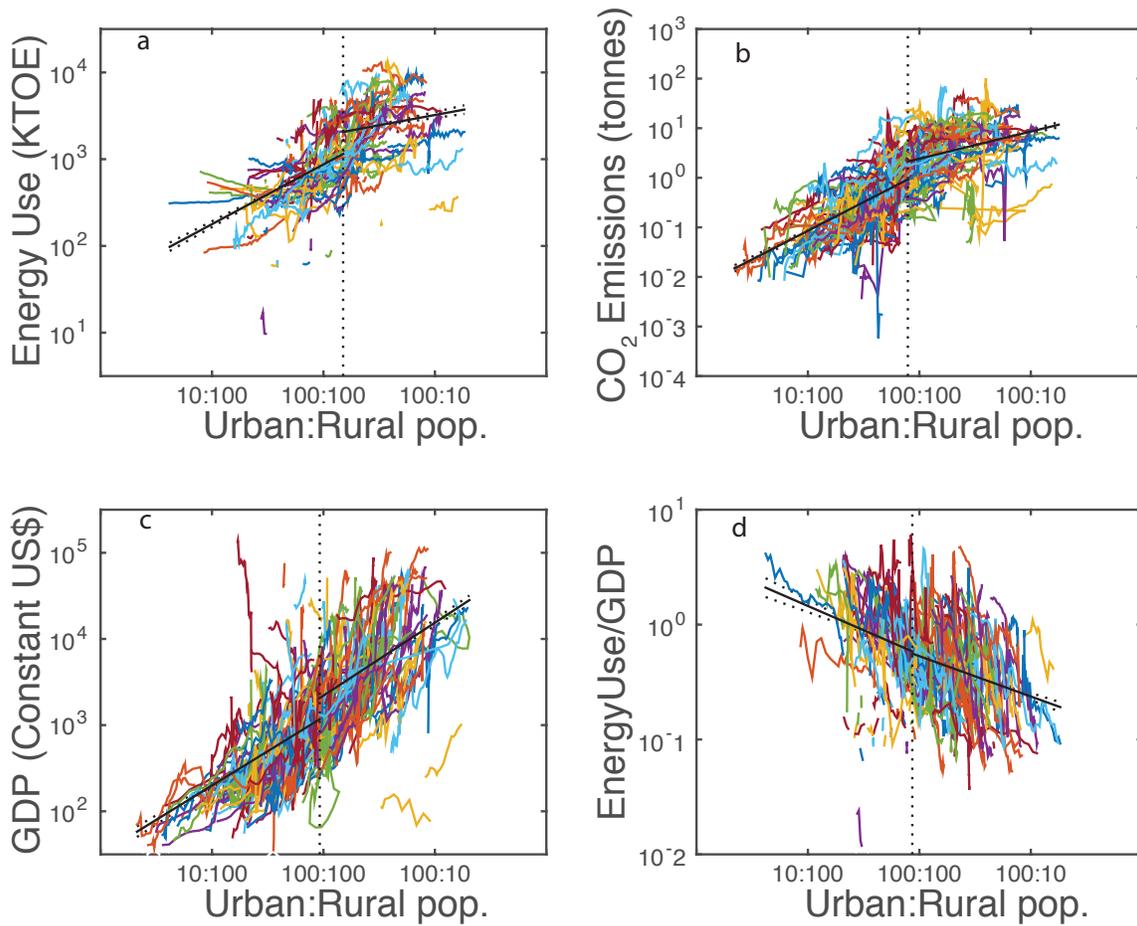

Figure 4: Log values of per capita energy use, $CO_2$ emissions, GDP, and energy use/GDP vs. urbanization (urban to rural ratio) across years and countries. Urban countries are to the right and rural countries to the left. Linear regressions (dark lines), and confidence intervals (broken lines) before and after change-points. Note that the dashed vertical lines represent the statistically determined breaks in the change-points, and not where urban/rural population is equal (e.g., 100:100). Linear fits to data before and after the change- points are in Table 1. Data are from the World Bank. See http://data.worldbank.org for most recently updated versions. Each colored line represents a given country overtime.

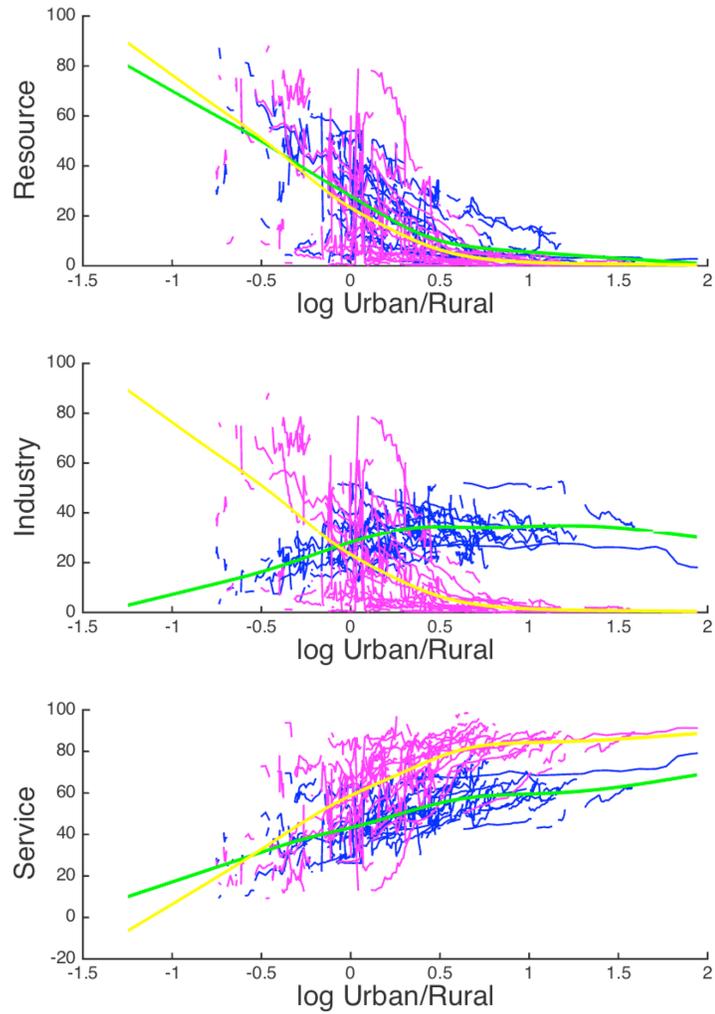

Figure 5: Percent employment in resource, industry, and service sectors of the economy by gender across the urban transition. Data are from http://data.worldbank.org. Each blue line represents male and magenta lines female employment for a given country over time. The green lines are spline fits of all male data and yellow lines are spline fits for all female data.

**Box 1: The urban transition in the United States.**

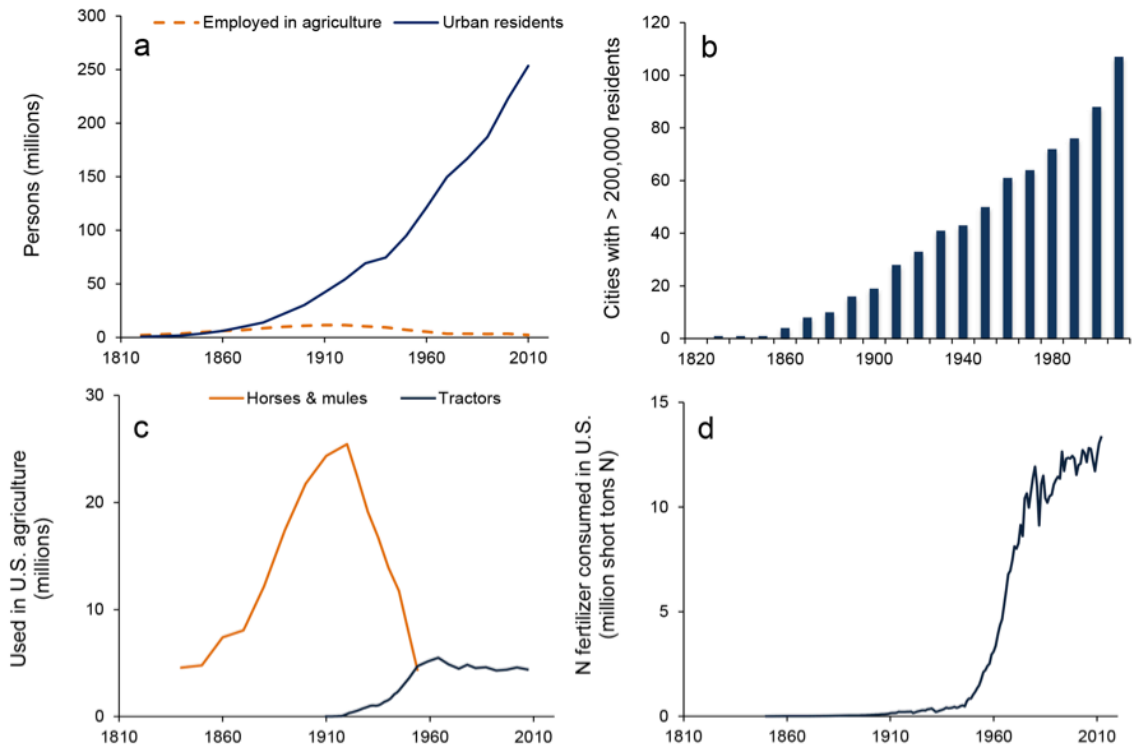

Figure 6: (a) U.S. persons employed in agriculture and residing in urban areas; (b) Number of U.S. cities with >200,000 residents; (c) horses/mules and tractors used in U.S. agriculture; (d) nitrogen fertilizer consumption in the U.S. Data sources: U.S. Census Bureau, The World Bank, United Nations (2011), USGS (2014), and FAOSTAT. Reprinted by permission from Springer Nature, America's Most Sustainable Cities and Regions by John Day and Charles Hall, Copyright 2016.

Data availability for the United States illustrates how resource use, food production, and population change as a country moves through the urban transition and develops an agricultural-industrial-technological economy. Draft animal labor, technological innovation, machinery, and cheap fossil fuels supported the growth of cities by reducing the number of human laborers required for agriculture (Figure 6, a). Today, >80% of the U.S. population resides in urban areas, while farmers, ranchers, and fishers currently contribute to less than 3% of employment in the U.S. and <2% is directly employed in agriculture (Vilsack and Clark, 2014, Figure 6, a). The total food-related share of the national U.S. energy budget is approximately 12.5% (https://www.ers.usda.gov/data-products/chart-gallery/gallery/chart-detail/?chartId=82504 (accessed: 17 April 2018) with more than 7 units of energy required to deliver 1 unit of edible food energy (CSS 2016, Canning et al. 2010, Heller and 2000; Aleklett, 2012). Camago et al. (2013) showed that thirteen major crops grown in the U.S. each require on the farm energy use between 6,922 and 21,235 MJ per hectare per year for production. Of this total production phase energy use, nitrogen fertilizer accounted for 36% on average, followed by on-farm fuel (30%), then potash fertilizers (7%), lime (6%), transportation of inputs (6%), phosphate fertilizers (5%), seed (5%), herbicide (4%), drying (2%), and insecticides (1%) (Camargo et al. 2013). Consumption of these resources grew rapidly during the 20$^{th}$ century alongside the rise of hyper-dense U.S. cities. For example, between 1961 and 1980, U.S. consumption of nitrogen fertilizer increased by 254%, slowing in recent years yet still exceeds 13 million tons per year (Fig 6, d; http://faostat.fao.org/). These inputs undoubtedly can enhance crop yields and thus conserve land in some cases. However, high input agriculture also continues to be energy-intensive, bear environmental burdens (Sinha et al. 2017), and often relies on finite resources subject to price volatility (Mew 2016).